\documentclass[preprint,aps,prb,showpacs]{revtex4}
\usepackage{amssymb,graphicx}

\begin{document}
\title{A Molecular Motor Constructed from
a Double-Walled Carbon Nanotube Driven by Axially Varying Voltage}

\author{Z. C. Tu}
\affiliation{Computational Materials Science Center, National
Institute for Materials Science, Tsukuba 305-0047, Japan}
\author{X. Hu}
\affiliation{Computational Materials Science Center, National
Institute for Materials Science, Tsukuba 305-0047,
Japan}
\begin{abstract}A new molecular motor is conceptually
constructed from a double-walled carbon nanotube (DWNT) consisting
of a long inner single-walled carbon nanotube (SWNT) and a short
outer SWNT with different chirality. The interaction between inner
and outer tubes is the sum of the Lennard-Jones potentials between
carbon atoms in inner tube and those in outer one. Within the
framework of Smoluchowski-Feynman ratchet, it is theoretically
shown that this system in an isothermal bath will exhibit a
unidirectional rotation in the presence of a varying axial
electrical voltage. Moreover, the possibility to manufacture this
electrical motor from DWNT is discussed under the current
conditions of experimental technique. 
\end{abstract}

\maketitle

In order to illustrate that the second law of thermodynamics
cannot be violated, Feynman introduced an imaginary
ratchet-and-paw system in his famous lectures. \cite{Feynman} This
system is now called Smoluchowski-Feynman ratchet because it was
first described by Smoluchowski in 1912. \cite{Smoluchowski} One
can observe the ratchet effect that the ratchet rotates
unidirectionally if two conditions are satisfied \cite{Reimann}:
(i) breaking of spatial inversion symmetry and (ii) breaking of
thermal equilibrium. Based on Smoluchowski-Feynman ratchet, many
statistical models are put forward, such as on-off
ratchets,\cite{Bug,Ajdari} fluctuating potential
ratchets,\cite{Astumian2} fluctuating force
ratchets,\cite{Magnasco,Luczka} temperature ratchets
\cite{Reimann2,Li,Bao} and so on. These models can explain the
directional matter transport at the molecular levels in biological
systems with the aid of the special protein
systems.\cite{Yanagida,Howard,Sambongi} Now the small devices or
systems at molecular levels are called molecular motors if they
can transform chemical, electrical or other forms of energy into
mechanical energy.

In another classic talk given by Feynman at the annual meeting of
the American Physical Society in 1959, he predicted that Human
would manufacture devices even at the molecular levels in the
future. \cite{Feynmanaps} His prediction are being realized with
the development of nanotechnology. \cite{Drexler} Especially,
after the discovery of carbon nanotubes, \cite{Iijima} researchers
designed many nanodevices, such as nanotube
gears,\cite{HanJ,Srivastava} nanotube
oscillators,\cite{ZhengQ1,Legoas,WangJW1,tzcosil} nanotube drills
\cite{Belikov,Lozovik} and so on. Tuzun predicted a doped nanotube
motor driven by laser, \cite{Tuzun} and one of the present authors
and Ou-Yang proposed a molecular motor constructed from a
double-walled carbon nanotube (DWNT) driven by temperature
variation. \cite{tzcmot1} Kang \textit{et al.} confirmed these
predictions by molecular dynamics simulations and then put forward
nanotube motors driven by fluid. \cite{KangJW2} Viewed from
statistical mechanics, the DWNT motor in Ref.~\onlinecite{tzcmot1}
is a kind of temperature ratchets \cite{Reimann2,Li,Bao} in
essence. But it is not easy to control the variation of
temperature. Thus we may ask: can we design a more convenient way,
for example using the electrical field, to drive this motor?

In this report, we show it is possible to construct a DWNT motor
driven by electrical field in physical point of view. The motor
consists of a long inner tube and a short outer tube. For
simplicity, the interaction between inner and outer tubes is taken
as the sum of the Lennard-Jones potentials between carbon atoms in
inner tube and those in outer one. Within the framework of
Smoluchowski-Feynman ratchet, we theoretically reveal that this
system in an isothermal bath will exhibit a unidirectional
rotation in the presence of a varying axial voltage. Moreover, we
also discuss the possibility to manufacture this electrical motor
under the current conditions of experimental technique.

\begin{figure}[!htp]
\includegraphics[width=10cm]{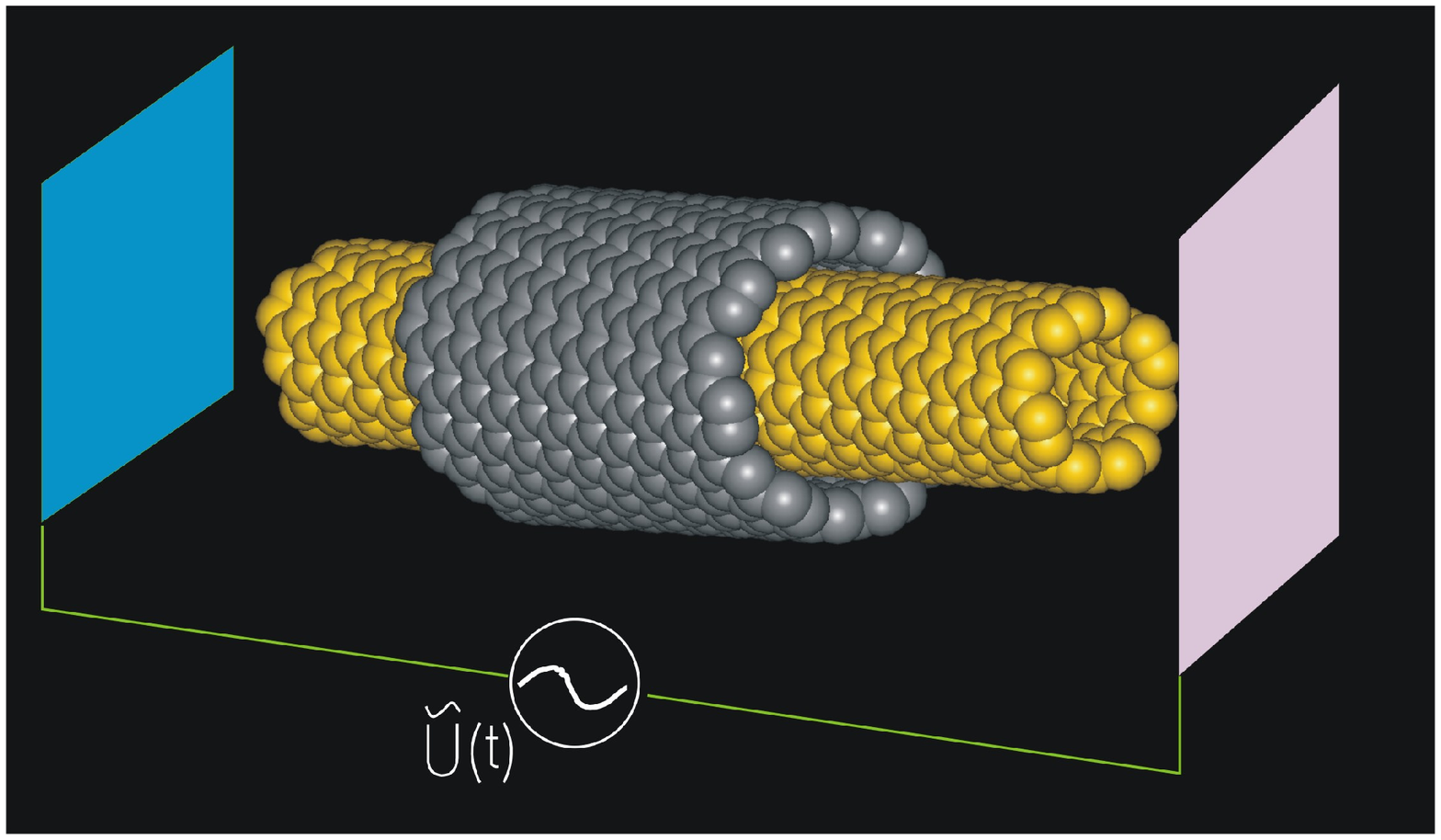}
\caption{\label{motorsch}Schematic diagram of the DWNT motor
driven by varying electrical field. The DWNT consists of a long
inner tube and a short outer tube. It is put between a pair of
electrodes with a varying voltage $\tilde{U}(t)$ applied on.}
\end{figure}

As illustrated in Fig.~\ref{motorsch}, the DWNT consists of a long
inner tube and a short outer tube with different chirality.
Without losing generality, we take (8, 4) and (14, 8) SWNTs, and
put two-unit cells for outer tube. The SWNTs have the same axis.
One end of inner tube is fixed while another is simply sustained.
We put the DWNT in between, without contacting to, a pair of
parallel electrodes on which we apply a voltage varying with time.
The axis of DWNT is perpendicular to the electrodes. The whole
system is put in a thermal bath, e.g., helium gas. Because of the
extremely high Young's modulus \cite{tzcyg} and stereo effect of
carbon nanotubes, the collisions of helium atoms at low
temperature, e.g., $T=50$ K almost have not effect on the shape of
carbon nanotubes as well as the coaxiality of two nanotubes. Two
degrees of freedom will be excited by the collisions: the rotation
of outer tube around the inner one and the slide of outer tube
along the axis. In following contents, we assume that some device
can fix the sliding degree of freedom of outer tube. The
interactions between atoms in outer tube and those in inner tube
provide potentials asymmetric with respect to the relative
rotation of the two nanotubes because of the difference in
chirality. This property breaks the spatial inversion symmetry and
naturally makes our device satisfy the first condition to exhibit
the ratchet effect.

The applied varying electrical voltage will achieve the second
condition of ratchet effect as discussed below. Recently Guo
\textit{et al.} found a large axial electrostrictive deformation
of SWNTs in the presence of axially electrical field by using
numerical Hartree-Fock and density functional calculations
\cite{Guowl} although SWNTs are usually regarded as
non-piezoelectric materials. According to these authors, the
electrical field changes the electronic structures of SWNTs and
induces elongations of carbon-carbon bonds without changing bond
angles. \cite{Guowl} Although their calculations are concerned to
the electrostatic field, it is easy to see that their results are
also available to the varying electrical field with period in the
order of nanoseconds because the response time of electron to
electrical field is usually about several femtoseconds. Thus, bond
elongations are synchronous with the variation of electrical
field, which induces the shaking interaction between outer and
inner tubes. Consequently, we obtain a fluctuating potential
which, according to Ref.~\onlinecite{Astumian2} is sufficient to
clear the second condition for ratchet effect in the statistical
point of view.

\begin{figure}[!htp]
\includegraphics[width=10cm]{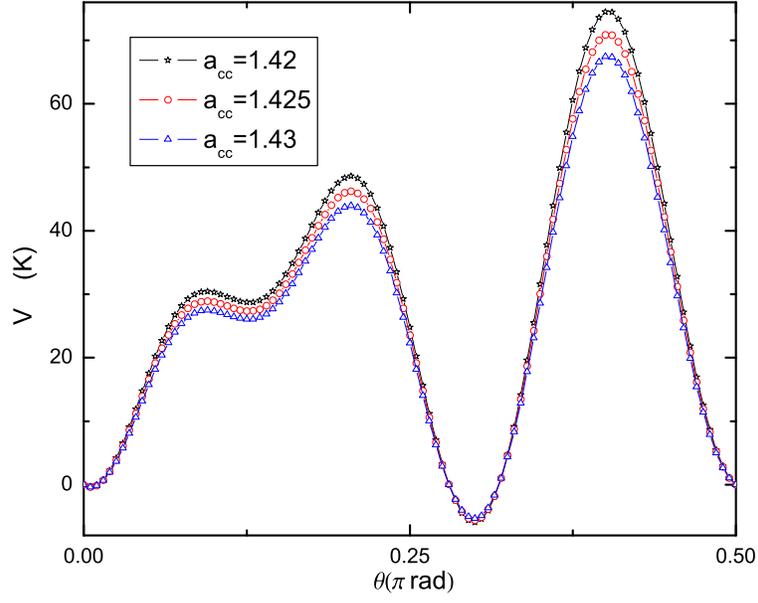}
\caption{\label{potent}Interactions between outer tube and inner
tube with different lengths of carbon-carbon bonds. Here the
potentials $V(0,a_{cc})$ are put to zero by subtracting some
constants.}
\end{figure}

To specify the above idea, we select an orthogonal coordinate
system $Oxyz$ by adopting the convention in
Ref.~\onlinecite{tzcpot} whose $z$-axis is the tube axis parallel
to the translation vector \cite{tzcpot} of SWNT, and $x$-axis
passes through some carbon atom in inner tube. The angle rotated
by outer tube around inner tube is denoted by $\theta$. All
coordinates of atoms are also calculated by the method in
Ref.~\onlinecite{tzcpot}. For theoretical simplicity, the
interaction between any atom $i$ in outer tube and atom $j$ in
inner tube is taken as Lennard-Jones potential \cite{lujp}
$u(r_{ij})=4\epsilon[(\sigma/r_{ij})^{12}-(\sigma/r_{ij})^6]$,
where $r_{ij}$ is the distance between atoms $i$ and $j$,
$\epsilon=28$ K, and $\sigma=3.4$\AA. \cite{hirschf} The inner
tube is long enough to be thought of as an infinite one. The total
interaction, $V(\theta, a_{cc})=\sum_{ij}u(r_{ij})$, between outer
tube and inner tube can be calculated by the method in
Ref.~\onlinecite{tzcpot}, where $a_{cc}$ is the length of
carbon-carbon bond whose value is $1.42$ \AA\ for zero voltage.
Obviously, the potential can be expanded by Taylor series
$V(\theta,a_{cc})\simeq V_0(\theta)[1+\alpha (\theta)\varepsilon]$
for small elongations of bond-length, where
$V_0(\theta)=V(\theta,1.42\mathrm{\AA})$, $\alpha (\theta)$ is a
function of $\theta$ and $\varepsilon$ the bond elongation ratio.
In Fig.~\ref{potent}, we show numerical results of the interaction
between outer tube and inner tube for different bond-lengths. Here
the potentials $V(0,a_{cc})$ are put to zero by subtracting some
constants which have no effect to our final results. We find that
the potentials are functions with period $\pi/2$ and that $\alpha
(\theta)$ is so weakly dependent on $\theta$ that we can fit it by
$\alpha (\theta)=-14.2$ from Fig.~\ref{potent}. On the other hand,
from Fig. 2 in the work of Guo \textit{et al.}, we know the bond
elongation ratio shows approximately a linear dependence on the
magnitude of electrical field with the slope $\beta=0.25$ \AA/V
provided that the applied voltage field is not too high. Moreover,
the field dependence of bond elongation ratio is insensitive to
the diameters and chiral angles of SWNTs. Therefore, we obtain a
shaking interaction, $V(\theta,t)=V_0(\theta)\{1+\mu[1+\cos(2\pi
t/\mathcal{T})]\}$, between outer tube and inner tube under a
varying voltage $\tilde{U}(t)=U_0[1+\cos(2\pi t/\mathcal{T})]$
between two electrodes, where $U_0$, $\mathcal{T}$ and
$\mu=\alpha\beta U_0/l$ are constant quantities. Here $l$ is the
distance between two electrodes and $\mathcal{T}$ is assumed
greater than 0.1 ns.

Next we will discuss the motion of outer tube in helium gas.
Because the mass of outer tube is much larger than that of helium
atom, its motion can be described by the Langevin equation $m
R^2\ddot{\theta}=-V'(\theta,t)-\eta
\dot{\theta}+\xi(t)$,\cite{Reichl} where $m\sim 10^{-23}$ Kg, the
mass of outer tube containing 496 carbon atoms, and $R=7.75$\AA\
is the radius of outer tube. $\eta$ is the rotating viscosity
coefficient, and the dot and the prime indicate, respectively,
differentiations with respect to time $t$ and angle $\theta$.
$\xi(t)$ is thermal noise which satisfies $\langle
\xi(t)\rangle=0$ and the fluctuation-dissipation relation
$\langle\xi(t)\xi(0)\rangle=2\eta T\delta(t)$, \cite{Reichl} where
$T=50$ K and the Boltzmann factor is set to 1. We estimate $m
R^2\ddot{\theta}/(\eta \dot{\theta})<10^{-3}$ by taking
$\ddot{\theta}\sim \dot{\theta}/\mathcal{T}$ and $\mathcal{T}>0.1$
ns. Thus it is reasonable to neglect the inertial term
$mR^2\ddot{\theta}$. In this case, the Fokker-Planck equation
corresponding to the Langevin equation is expressed as
\cite{Reimann}
\begin{equation}\label{fokkerpl}
\frac{\partial P(\theta ,t)}{\partial t} =\frac{\partial
}{\partial \theta }\left[ \frac{V^{\prime }(\theta,t )}{\eta
}P(\theta ,t)\right] +\frac{T}{\eta }\frac{\partial ^{2}P(\theta
,t)}{\partial \theta ^{2}},
\end{equation}
where $P(\theta,t)$ represents the probability of finding the
outer tube at angle $\theta$ and time $t$ which satisfies
$P(\theta+\pi/2,t)=P(\theta,t)$ and
$\int_0^{\pi/2}P(\theta,t)d\theta=1$. It follows that the average
angular velocity of outer tube in the long-time limit
\cite{Reimann}
\begin{equation}\label{current}
\langle\dot{\theta}\rangle=\lim_{t\rightarrow
\infty}\frac{1}{\mathcal{T}}\int_t^{t+\mathcal{T}}dt
\int_0^{\pi/2}d\theta\left[-\frac{V'(\theta,t)P(\theta,t)}{\eta}\right].
\end{equation}

Let $D=\frac{T}{\eta}$, $t=D^{-1}\tau$,
$\mathcal{T}=D^{-1}\mathcal{J}$, $V_0(\theta)=u(\theta)T$, we
obtain the dimensionless forms of (\ref{fokkerpl}) and
(\ref{current}):{\small
\begin{eqnarray}
&&\frac{\partial P(\theta ,\tau )}{\partial \tau } =\frac{\partial }{%
\partial \theta }\left\{ \left[ 1+\mu \left(1+\cos \frac{2\pi \tau}
{\mathcal{J}}\right)\right]u^{\prime }(\theta ) P(\theta ,\tau
)\right\} +\frac{\partial ^{2}P(\theta
,\tau )}{\partial \theta ^{2}}, \label{fokkerdim}\\
&&\left\langle \frac{d\theta }{d\tau }\right\rangle =-\lim_{\tau \rightarrow \infty }%
\frac{1}{\mathcal{J}}\int_{\tau }^{\tau +\mathcal{J}}d\tau
\int_{0}^{\pi /2}d\theta \left[ 1+\mu \left(1+\cos \frac{2\pi
\tau} {\mathcal{J}}\right) \right]u^{\prime }(\theta ) P(\theta
,\tau ).\label{currentdim}
\end{eqnarray}}

Because $V_0(\theta)$ is a periodic function, we can expand it by
Fourier series and find that it is well fit by
$V_0(\theta)=29.12-0.18\cos 4\theta-16.14\cos 8\theta-12.53 \cos
12\theta-4.48 \sin 4\theta-19.52 \sin 8\theta+11.52\sin 12\theta$
(K), where high-order terms are neglected because their
coefficients are very small. Thus
$u'(\theta)=V_0'(\theta)/T=\sum_{k=1}^3(v_{k}e^{4i k\theta
}+v_{k}^{\ast }e^{-4ik\theta })$ with $v_1=-0.179 - 0.007i$,
$v_2=-1.562 - 1.291i$ and $v_3=1.382 - 1.504i$. In the the
long-time limit, $P(\theta,\tau)$ can be expanded by Fourier
series $P(\theta ,\tau )=\sum_{n,m=-\infty}^{\infty} p_{nm}\exp
i[\left( 2n\pi \tau /\mathcal{J}\right) +4m\theta ]$. Substituting
them into (\ref{fokkerdim}), we arrive at a recursion equation
\begin{equation}\label{recusion}
p_{nm} =\frac{2im\mathcal{R}_{nm}}{8m^{2}+i\left( n\pi \tau
/\mathcal{J}\right) },
\end{equation}
with $\mathcal{R}_{nm}=\sum_{k=1}^3\{(1+\mu)
(v_kp_{n,m-k}+v_k^*p_{n,m+k})
+(\mu/2)[v_k(p_{n-1,m-k}+p_{n+1,m-k})
+v_k^*(p_{n-1,m+k}+p_{n+1,m+k})]\}$. Similarly, (\ref{currentdim})
is transformed into
\begin{equation}\left\langle \frac{d\theta
}{d\tau }\right\rangle =-(\pi
/2)\mathcal{R}_{00}.\label{fluxrs}\end{equation}

\begin{figure}[!htp]
\includegraphics[width=10cm]{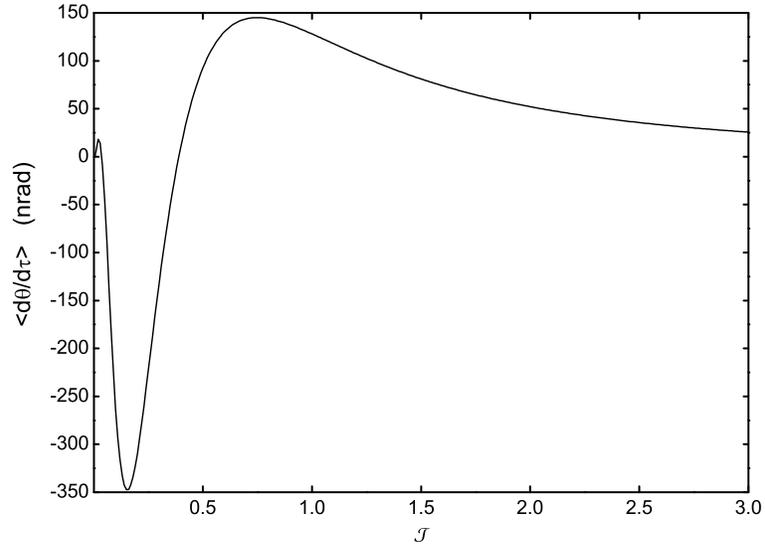}
\caption{\label{flux}Average dimensionless angular velocity
$\langle d\theta/d\tau\rangle$ of outer tube rotating around inner
one in isothermal bath when varying axial voltage with
dimensionless period $\mathcal{J}$ is applied. The minus sign
means the clockwise rotation around $z$-axis while the positive
represents counterclockwise rotation.}
\end{figure}

Considering the constraint $p_{n0}=(2/\pi )\delta _{n0}$ coming
from $\int_0^{\pi/2}P(\theta,t)d\theta=1$, we solve the recursion
equation (\ref{recusion}) with $\mu=-0.01$ for different
dimensionless period $\mathcal{J}$, and then calculate the
dimensionless average angular velocity $\left\langle \frac{d\theta
}{d\tau }\right\rangle$ by (\ref{fluxrs}). The dimensionless
period dependence of average angular velocity of outer tube
rotating around inner tube is shown in Fig.~\ref{flux}, from which
we find that: (i) $\left\langle \frac{d\theta }{d\tau
}\right\rangle\rightarrow 0$ if $\mathcal{J}\rightarrow \infty$
which is equivalent to the case of an electrostatic voltage; (ii)
$\left\langle \frac{d\theta }{d\tau }\right\rangle=0$ if
$\mathcal{J}=0$ which corresponds to the case that the voltage
varies so quickly that outer tube can not respond to it in time;
(iii) Outer tube rotates counterclockwise around the tube axis for
some $\mathcal{J}$, which corresponds to the positive value of
$\left\langle \frac{d\theta }{d\tau }\right\rangle$, and clockwise
for other $\mathcal{J}$, which corresponds to the minus value of
$\left\langle \frac{d\theta }{d\tau }\right\rangle$; (iv)
$\left\langle \frac{d\theta }{d\tau }\right\rangle$ has a
significant value $-347$ nrad when $\mathcal{J}_c=0.15$.

For helium at $T=50$ K, we can calculate $\eta=1722$ K$\cdot$ns
from its value at 273 K. \cite{Nordling} It is necessary to
emphasize that $\eta$ is the rotating viscosity coefficient which
is different from the viscosity coefficient of gas $\eta_c$ in the
common sense. The simple relationship between them is $\eta=2\pi
R^2 L \eta_c$ for cylinders, where $L=27.39$ \AA\ is the length of
outer tube. Therefore we obtain
$\langle\dot{\theta}\rangle=-10.08$ nrad/ns (i.e., about one and a
half rounds per second) when the period $\mathcal{T}_c=5.17$ ns.
Here we obtain these values in disregard of the fiction between
nanotubes because it is very small. \cite{Cumings}

The above discussions demonstrate that we can construct an
electrical motor from DWNT in principle. Two key points are: (i)
DWNT with different outer and inner chirality induces a potential
that breaks spatial inversion symmetry; (ii) Some mechanism
(varying voltage in this paper) induces the shaking of the
potential that breaks thermal equilibrium. Therefore, similar
results are available in any DWNT with different outer and inner
chirality although our discussions are focused on (8,4)@(14,8)
tube. Additionally, we use the results obtained from density
functional theory in Ref.~\onlinecite{Guowl} that shows large
axial electrostrictive deformation for armchair and zigzag tubes
with different diamters. These investigated cases reveal that the
electrostrictive deformation is insensitive to the diameters and
chiral angles of SWNTs. Generally speaking, different voltage
dependence of bond-elongation might exist for different chirality.
Even if were the case, it does not hinder from inducing the
shaking of the potential. In short, different DWNT and different
voltage dependence of bond-elongation do not change the
qualitative conclusion of this paper but merely modify the
quantitive value of the average angular velocity.

Can this device be made in reality? There are several
technological difficulties. The first one is to make the DWNT with
different chirality of outer and inner tubes. Experimentalists can
overcome it by producing DWNT based on the synthesis technique in
Ref. \onlinecite{Cumings2}. The second difficulty is the varying
voltage.
 Our device requires voltage ($U_0\sim3$ V for
$l\sim 0.1$ $\mu$m) with high frequency of gigahertz. The third
one is the method to restrict the sliding degree of freedom of
outer tube. It is possible to manufacture the DWNT motor driven by
varying electrical voltage provided that researchers overcome the
above difficulties.

In summary, we conceptually design a new molecular motor from DWNT
driven by varying electrical voltage. In this sense, it can be
called ``electrical motor" in nanometer scale. Similarly, we may
design another molecular motor if we replace carbon nanotubes with
boron nitride nanotubes because the latter is a piezoelectric
material. \cite{Nakhmanson} People have a long dream to make
machines in nanoscales. The key component to these machines is the
power device. Our motor is expected to be the power device if it
is manufactured in the future with the development of
nanotechnology.

We are grateful to Dr. Q. X. Li and B. Y. Zhu for their technical
helps.

\newpage

\newpage
\newpage
\end{document}